\documentclass[12pt]{article}
\usepackage{amsfonts,epsfig,latexsym,amscd}

\begin{document}
\begin{titlepage}
\title{
\vskip -70pt
\begin{flushright}
{\normalsize \ DAMTP-2002-144}\\
{\normalsize \ physics/0211069}\\
\end{flushright}
\vskip 20pt
{\bf An Fe-Si-Ni solidification model of the Earth's layering}
}
\vspace{1cm}
\author{{A. Aitta}\thanks{e-mail address: A.Aitta@damtp.cam.ac.uk}\\
{\sl Department of Applied Mathematics and Theoretical Physics}\\
{\sl University of Cambridge} \\
{\sl Silver Street, Cambridge CB3 9EW, U.K.}}
\date{Nov 15, 2002}
\maketitle
\thispagestyle{empty}
\vspace{1cm}
\begin{abstract}
\noindent
The physical process creating layered structure in planetary rocky
bodies is considered here to be multicomponent solidification. This is a
unified alternative approach to the present interpretations where each
layer is reasoned and matched individually. The Earth's solidification
is modelled using the ternary phase diagram
Fe-Si-Ni. The four cotectic concentrations and the four corresponding
seismic discontinuity radii have been used to show that the silicon
concentration as a function of the distance $R$ from the centre of the
Earth can be modelled by $C_{Si} = \Gamma (R/R_E)^2$ where $\Gamma$ =
0.583,  $R_E$  is the
Earth's radius and the Ni/Fe ratio is 0.072.  Earth would have up to
13 chemically different layers. This model predicts that there are three
to four chemically slightly different sublayers in the D" layer and  two
boundaries in the inner core  at radii 870 km and 1050 km. The
observed hemispheric asymmetry in the inner core could follow if the
Ni-concentration slightly varies locally. Although this model can
account for all the layers using only three elements it can not, of course,
match the full chemistry of the real Earth.

\end{abstract}
\end{titlepage}

\section{INTRODUCTION}
\label{sec:intro}

Whether terrestrial objects in their early stages were in a completely
molten state or not may be a key determinant for their final state
\cite{Ag}. Here the observed structure of the Earth is considered as
have been formed by the solidification process of an initially liquid
Earth. Gravity acting with thermal convection is assumed to have
quadratically stratified it before the solidification began, rather as
the liquid outer core now has a density depending quadratically on the
radial distance \cite{DA}.
 
The main features of the solidification process occurring in a
multicomponent real planet can be modeled using a much simpler fluid
system having only a few components. In this paper a ternary system
\cite{Ai2, RR} has been used. By having
the third component one can reach more realistic conclusions than has
been possible previously using a binary solidification model
\cite{Br}. On the other hand, a more than three component
system would be much harder to analyze and compare reliably with the
data.

The rocky planets' main two components are silicon and iron. Due to
gravity, the lighter Si is predominantly in the mantle and the heavier
Fe in the core. Based on meteorite findings it is commonly believed
that there is also nickel in the core. This paper considers structure
formation in an Earth-size planet made only of three components,
namely Fe, Si and Ni. Ni occurs only in a small quantity, but is
important in the inner core where it is the most abundant element after
Fe. Being lightest, the concentration of Si increases from the centre,
while both Fe and Ni have their maximum concentration there. There is
no assumption of a complete segregation into a silicon mantle and iron
core. However, the assumed quadratic increase of the Si concentration
with radius, and similarly decreasing Fe and Ni concentrations, allow
the liquid in the outer core to be much enriched in Fe and depleted in
Si. Thus Si represents here the light element(s) needed to be present
in the outer core to produce the seismically deduced density which is
6-10 \% less than that estimated for pure liquid Fe. The identity of
the light component(s) has been discussed
for 50 years, the discussion being initiated by Birch \cite{Bi}, and all this time Si has 
been included among the leading candidates \cite{Po}.

When the planet's surface has cooled enough through thermal radiation
into space, surface solidification of Si occurs. The quadratic density
profile allows the Si solid crystals to float rather than sink as
might occur with a homogeneous density profile. So the solidification
front progresses inwards from the surface. The high pressure, despite
being combined with high temperature in the centre of the planet,
initiates another solidification front starting from the centre and
progressing outwards.

\begin{figure}
\center
\epsfig{file=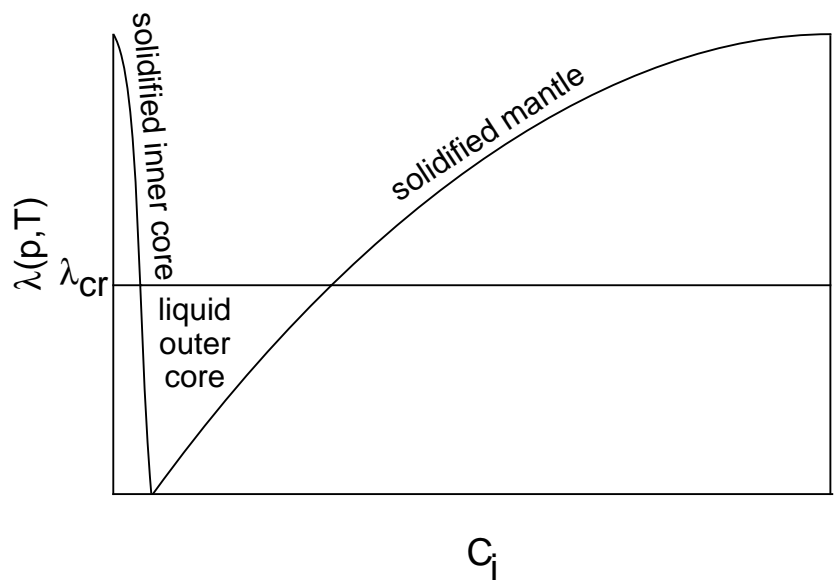,width=12cm}
\flushleft
\caption{
{\small \label{Fig. 1.}
A schematic presentation of a planetary solidification
process. The 
vertical axis indicates some solidification parameter $\lambda$ which
depends on both 
pressure $p$ and temperature $T$ and $C_i$ is the concentration of the
lightest 
element. A critical value $\lambda_{cr}$ separates the solidified
phases from the liquid 
phase. 
}}
\end{figure}

These solidification fronts are schematically presented in Fig. 1
where the horizontal axis represents the concentration of the lightest
element and the vertical axis represents some solidification
parameter $\lambda$ which depends on both temperature and pressure. Fluid
exists only for $\lambda < \lambda_{cr}$, which in the figure is assumed
to be the same for both fronts. The details of the solidification
product can be seen 
from the Fe-Si-Ni ternary phase diagram.

\section{TERNARY PHASE DIAGRAM OF FE-SI-NI AND THE SOLIDIFICATION PROCESS}
\setcounter{equation}{0}
\label{sec:svfm}

Fig. 2 shows the relevant part of the ternary phase diagram of Fe, Si and Ni
at normal pressures \cite{RR}, represented here as
having the Fe-corner a  right angle. Above its horizontal axis is
sketched the Fe-Si binary phase diagram with its temperature
dependence shown vertically \cite{RR, ASM}. It has several critical
points on its liquidus surface: between
them the crystals precipitating from the liquid would be made of the
compound of Fe and Si indicated on the graph, but cocrystallization of
the crystals of both neighbouring regimes occurs at the critical
points. The Ni component does not influence the concentrations where
the Fe-Si binary system has its phase boundaries: they would stay the
same whatever the third component was. (See, for instance, the phase
diagrams for Fe-Si-O, Fe-Si-Al \cite{ASM}). All the eutectic
type critical points continue as lines inside the ternary phase
diagram. Two peritectic critical points $P_1$ and $P_2$ are the least well
agreed upon \cite{RR} and their continuations are
omitted although they  should be there, too, but they do not have any
role for the Earth since their concentrations correspond to the still
liquid Earth. Otherwise the liquidus line is reliably established from
Fe to Si \cite{RR}.

\begin{figure}
\vskip -2.3cm
\center
\epsfig{file=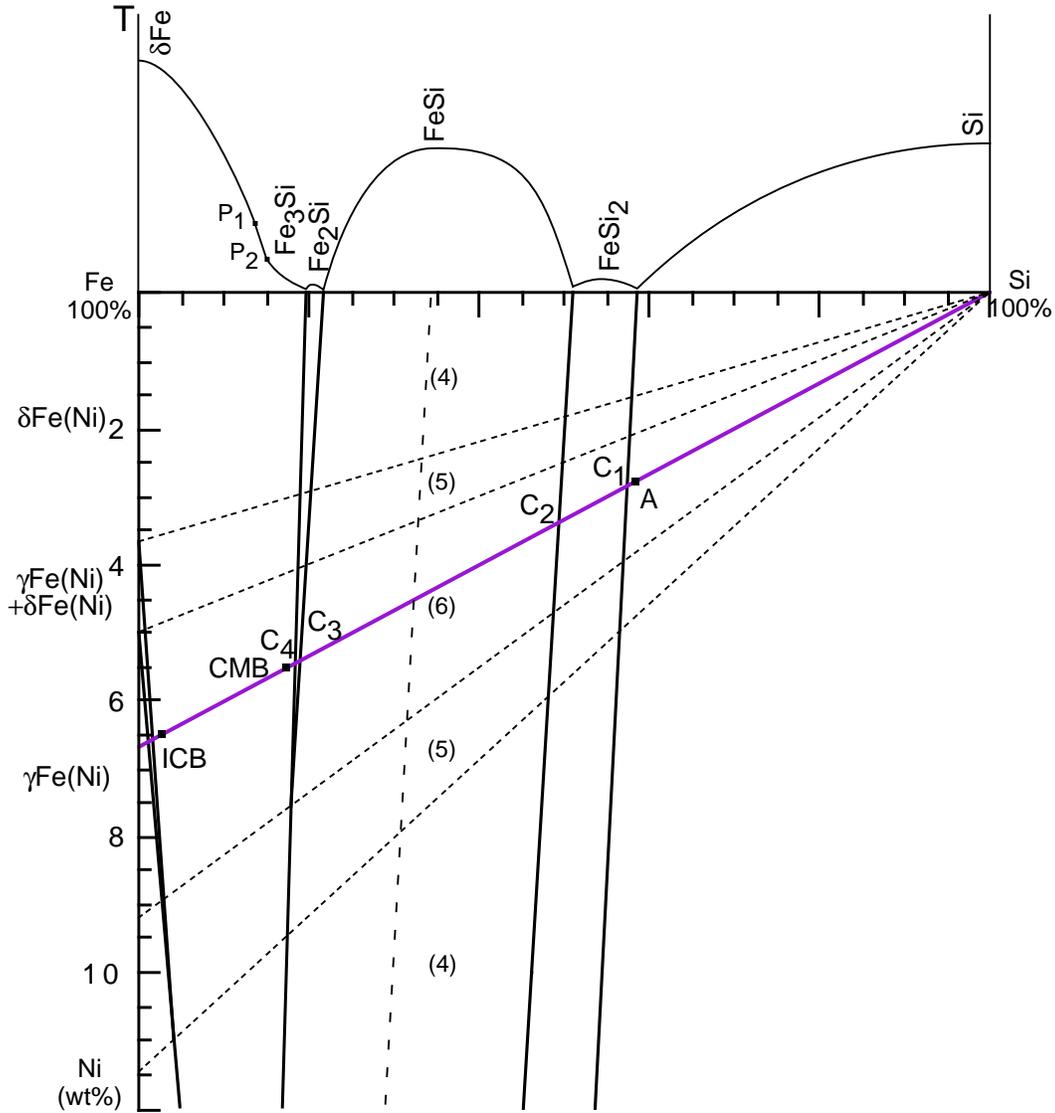,width=14cm}
\flushleft
\vskip -.5cm
\caption{{\footnotesize
\label{Fig. 2.} Fe-rich corner of the Fe-Si-Ni ternary phase diagram
 \cite{RR, ASM}. Above the horizontal axis the liquidus surface of the
 Fe-Si binary phase  
diagram  \cite{RR, ASM} has been sketched with its four eutectic
points and its different 
crystallization products indicated. The eutectic points continue as cotectic 
lines into the interior of the phase diagram. The Fe-Ni axis has a narrow 
peritectic region, which has been linearly continued towards the point 
(Si, Ni)=(0.04, 0.11) where it changes to be a cotectic line. The residual fluid 
concentration  evolves from A (the initial fluid concentration) along
the tie line  
towards $C_1$ as the Si-mush is solidifying. At $C_1$ a cotectic
mixture of Si and  
FeSi$_2$ is solidifying. After $C_1$ the residual fluid evolves along
the same tie line  
towards $C_2, C_3, C_4$ and on to the CMB, solidifying between them a mushy 
compound and on each $C_j$ a cotectic mixture of the neighbouring phases. The 
inner core has been assumed to have started to solidify from the other end of 
this same line and the residual fluid concentration on its
solidification front  
has now crossed the peritectic region and has reached the ICB (inner core 
boundary). The liquid line of descent through A corresponds to an Ni/(Ni+Fe) 
ratio of 0.067. For different planetary bodies the slope of this line
(Ni/(Ni+Fe)  
ratio) could vary and the five possible regions are separated by dotted lines 
with different selections of the crossed phase boundaries, the number of 
them being shown in brackets. The dashed line approximates where the solid 
fraction in the FeSi phase reaches its maximum.
}}
\end{figure}

The vertical edge, the Fe-Ni binary system, needs to be considered
here only for small Ni-concentrations. For $k$ = Ni/(Ni+Fe) $\le 0.037$
there is an Ni-poor $\delta$Fe(Ni) phase and for $k \ge$ 0.05 an Ni-rich
$\gamma$Fe(Ni)
phase. Between them there is a peritectic regime where both phases
occur. These phase boundaries are extended as straight lines into the
interior of the phase diagram and they join at (Si, Ni) = (0.04, 0.11)
where the peritectic type transition changes to eutectic type
\cite{RR, ASM}.

In Fig. 2,  the axis of the critical solidification parameter $\lambda$ should
be visualized as perpendicular to the concentration plane similarly as
in the simpler Fig. 1. $\lambda$ depends on both temperature and pressure,
combined in such a way that the solidification process, once started,
continues: the thermal barriers present in  phase  diagrams for normal
pressures are not obstacles here since they can be overcome by the
changing pressure. Thus the thermal barriers do not prevent the liquid
line of descent being approximated by a straight line throughout the
ternary phase diagram. The liquid planet's initial surface
concentration is represented by the point A in Fig. 2. When the liquid
is cool enough, Si starts to crystallize out, forming mush: porous
solid having residual liquid in its interstices. The released fluid is
enriched in Fe and Ni. If there were no convection \cite{Ai1, Ai2},
the residual liquid concentration at
the solidification front would follow the linear path moving away from
the Si-corner. There the ratio of the Ni to Fe concentration is
constant. However, there is both thermal and compositional convection
\cite{Ai2} which carry the heavy elements (whose density is
similar) down and lighter Si up, holding the concentrations suitably
balanced at the front so that the linear path of descent is still
followed. When the liquid concentration reaches the first phase
boundary at $C_1$ in Fig. 2, a cotectic crystallization of Si and FeSi$_2$
starts, firstly inside the earlier formed Si-mush, but finally as a
new front on its own reaching deeper down than the Si-mush. This
cotectic product is also a mush, but its solid fraction is greater
than in the Si-mush. The non-solidifying element is Ni, which is the
heaviest, and would, if in excess, descend under gravity. Thus the
liquid line of descent does not turn along the cotectic line as it
would do if there were no convection \cite{Ai1, Ai2}. Cotectic
solidification of Si and FeSi$_2$ continues at $C_1$ as
long as there is enough available Si to crystallize both of
them. After that only FeSi$_2$ crystallizes and the residual liquid
concentration continues to increase in Fe and Ni along the line
$C_1C_2$. When the concentration reaches the next cotectic line at $C_2$,
cocrystallization of FeSi$_2$ and FeSi occurs as long as there is
adequate Si available. When not, the phase FeSi would crystallize
alone, releasing only Ni, which descends below the solidification
front. The liquid line of descent continues its straight path. The
next cotectic line induces at $C_3$ cocrystallization of FeSi and Fe$_2$Si,
consuming all excess Si. A pure Fe$_2$Si phase follows as the
solidification front moves further down towards the planet's
centre. The next crystallization product is the cotectic mixture of
Fe$_2$Si and Fe$_3$Si at C$_4$, followed by pure Fe$_3$Si mush as far as CMB
(core-mantle boundary). 

For simplicity, the Ni/Fe ratio has been assumed constant along the
liquid line of descent, even near the Fe-Ni edge.  Any variation
induced by the two different solid solutions of Fe(Ni) or by
convection is expected to be at most very small. The iron-rich end of the line
represents the fluid concentration in the centre of the planet. There
the crystallization forms a solid solution of Fe(Ni). When Si is
present in small quantities, the solidification of Fe(Ni) increases
the amount of Si in the fluid, but since it is the lightest component,
it rises. However, a small amount of this residual liquid is trapped
in the interstices of the mush. If $k \ge 0.05$  the first solid to form
has an Ni-rich $\gamma$Fe(Ni)-structure. When the amounts of Fe and Ni have
been consumed sufficiently, there would be simultaneous
crystallization of both Ni-rich $ \gamma$Fe(Ni) and  Ni-poor
$\delta$Fe(Ni).  Further
solidification of Fe and Ni leads to the phase of Ni-poor $\delta$Fe(Ni), in
the form of mush. If the planet were completely solidified, this
outward growing phase would meet the downward growing Fe$_3$Si phase.

\section{QUANTITATIVE RESULTS CORRESPONDING TO THE EARTH}
\label{sec:psf}
\setcounter{equation}{0}

The fluid density in the liquid outer core depends in the PREM model
\cite{DA} quadratically on radius, and this has
motivated the assumption here that the concentration of the initial
fluid depends on the radial variable quadratically: Si is increasing
from the centre, Fe and Ni are decreasing from their maximum value at
the centre to some non-zero value at the surface. The precise
quadratic dependence can be inferred from the seismic data as
explained below.

All phase boundaries in Fig. 2 are straight lines $C_{Ni} = k_j C_{Si} + b_j$,
which are valid for the small values of $C_{Ni}$ considered here, and the
values for $k_j$ and $b_j$ for each phase boundary were obtained from the
graph in \cite{RR} for the four Si-rich lines and from
\cite{RR, ASM} 
for the two Si-poor
lines. The liquid line of descent $C_{Ni}  =  k (1-C_{Si})$ crosses these
phase boundaries in 4 points if $k$ is less than 0.037, in 5 points if
0.037 $\le k <$ 0.05, in 6 points if $0.05 \le k < 0.092$ and again 5 points
for  $0.092  \le  k < 0.114$. For bigger $k$ there are only 4 crossing
points. These regimes are separated by dotted lines in Fig. 2. The $C_{Si}$
values of these crossing points were sought by changing the slope of
the liquid line of descent in steps of 0.001. Their corresponding 
$R$-values were estimated as discussed below. The coefficient $a$ in the
relationship $C_{Si} = a R^2$, together with $k$, could then be found by a best fit.

The density discontinuities in the Earth according to the PREM model
are presented in Fig. 3. The discontinuities have been identified with
the phase changes in the ternary phase diagram. Fig. 3a shows the top
800 km. The outermost layer, crust, varies from 6 km under the oceans
to 80 km below the continents, and is shown shaded. In the
three-component Earth model the crust has been identified as made of
Si crystals. The Si concentration on the cotectic line Si+FeSi$_2$ has
been used paired with both $R$ = 6365 km (appropriate for oceans) and $R$
= 6291 km (appropriate for continents) allowing the fit to find the
most  suitable thickness for the Si layer. The small density jump seen
in PREM at 220 km depth is now known to occur only locally and is not
used.  The 400 km depth density jump separating the upper mantle from
the transition zone has been interpreted as corresponding to the end
of the cocrystallization of Si and FeSi$_2$ and to the beginning of the
pure FeSi$_2$ phase. This depth is not used in the fit since only the
initiation of the cocrystallization concentrations are available in
the phase diagram. The next seismic discontinuity, which is much
weaker and not visible in PREM, occurs at about 520 km depth, and has
been seen in some regions to split into a pair of discontinuities at
about 500 km and 560 km \cite{DW}, also shown shaded
in Fig. 3a. This has been identified as the beginning of the cotectic
layer of FeSi$_2$+FeSi indicated by the second cotectic line from the
right in the phase diagram. All three $R$-values were used in the fit,
with the corresponding Si concentration on this phase boundary, to
allow the fit sufficient freedom to address the radial
uncertainty. This cotectic phase would continue until the end of the
transition zone. The density jump at 670 km depth has been interpreted
as the beginning of the pure FeSi phase which makes up the over 2000
km thick lower mantle, down to the D" layer.

\begin{figure}
\vskip -2.2cm
\center
\epsfig{file=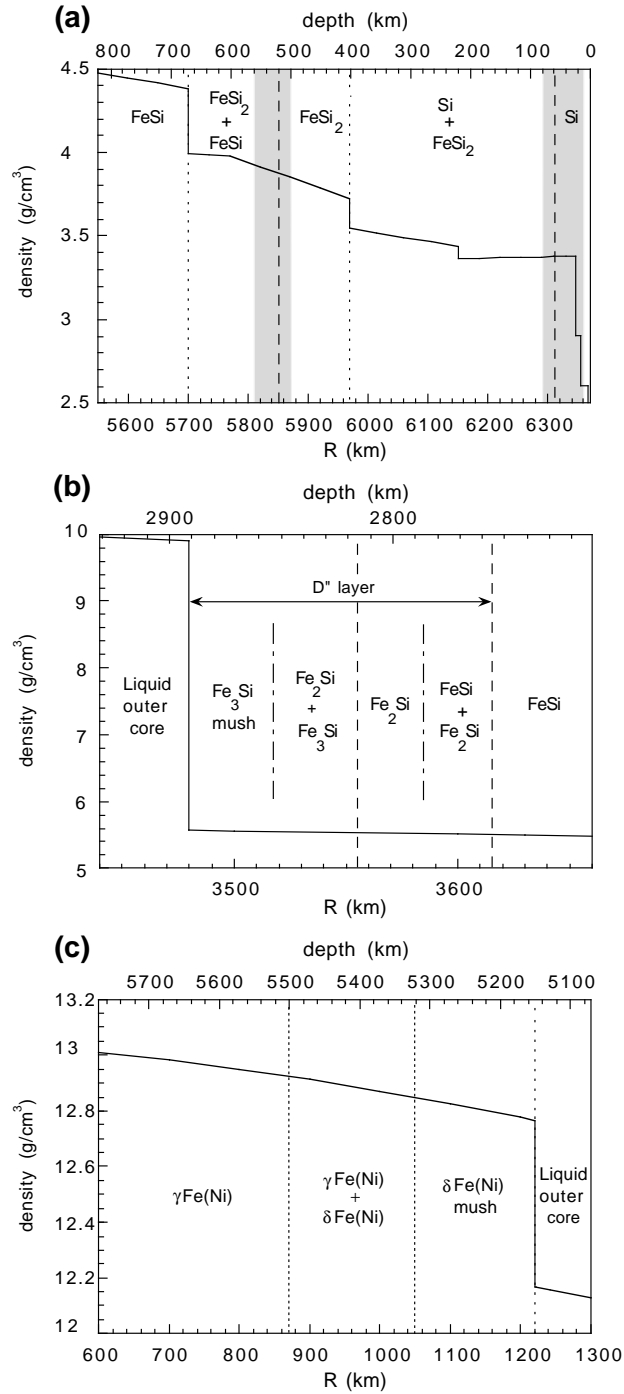,width=8.5cm}
\flushleft
\vskip -.8cm
\caption{{\footnotesize
\label{Fig. 3.}
 Earth's layered structure and its interpretation using the ternary phase 
diagram, drawn on the top of the density profile according to the PREM model 
\cite{DA}. (a)  The top 800 km of the Earth. Shaded areas indicate the
radial regions 
used by the fit to find a suitable Moho radius and the smaller seismic 
discontinuity occurring around 520 km depth. The dashed lines are the 
results from the fit. The bigger density discontinuities, marked as
dotted lines,  
have been interpreted to be the end radii of the cotectic crystallization. (b) 
Region at the bottom of the mantle. PREM indicates only the CMB density 
jump. The D" layer was identified to consist of at least the Fe$_2$Si
phase with its  
cotectic neighbouring phases. Its lower radius was sought within 100 km of 
the CMB and its upper radius in the next 100 km above. The dashed lines are 
the results from the fit. The lower radii of the cotectic layers can not be 
determined by the model, the dashed-dotted lines are just indications that 
those boundaries should be somewhere there. (c) The inner core structure 
predicted by this model. 
}}
\end{figure}

The core-mantle boundary (CMB) region is shown in Fig. 3b. It gives
(for $k < 0.092$) two further restrictions for $C_{Si}(R)$. Since in the
lowest mantle there is a special, seismically distinct D" layer whose
thickness is uncertain but varies from about 100 km up to perhaps 300
km in some places \cite{MS}, the following
interpretation has been made. The CMB was used to estimate where the
cotectic crystallization of Fe$_2$Si and Fe$_3$Si occurs and the fit was
guided by this Si concentration combined with $R$-values $R_{CMB}$ = 3483
km \cite{MS} and $R_{CMB}$ + 100 km to allow some necessary
flexibility in its exact location. The upper edge of the D" layer was
sought in the region of the next 100 km (using $R$-values of 3583 and
3683 km) it being an adequate interval for the data used.  It was
interpreted to correspond to the beginning of the cotectic
crystallization of FeSi+Fe$_2$Si. These values guided the fit in this
region.

The best fit, with the biggest correlation coefficient and the
smallest $\chi^2$, was obtained using the Si-concentrations of these four
phase boundaries together with the $R$-values obtained from the
seismic and/or density discontinuities, calculated for each slope $k$
(Ni/(Ni+Fe) ratio). The result is that in the fluid $C_{Si} = a R^2$
where $a = (1.437 \pm 0.007) . 10^{-8}$ km$^{-2}$, and $k$=
0.067. This, if scaled by the 
Earth's radius $R_E$ = 6371 km, gives $C_{Si} = \Gamma (R/R_E)^2$ with   
$\Gamma$ = 0.583
which determines the Si-concentration at point A in Fig. 2. The
corresponding Ni/Fe ratio is 0.072. The fit, with the points guiding
it, is presented in Fig. 4. Also shown is the radial dependence of the
other components Ni and Fe, and the radial locations of the phase
boundaries and their chemical structure. Using the Si-concentration
formula and the Si-concentrations on the two remaining lines in the
phase diagram, with the same $k$, one obtains predictions for two
distinct radii (see Fig. 3c and solid dots in Fig. 4) in the inner
core where the chemical content and the crystal structure change. The
PREM model does not include this possibility. The innermost boundary
corresponds to $R$  = 870 km and the outer one to $R$ = 1050 km.  This
allows one to interpret the seismically isotropic top 100 - 200 km
thick  layer of the inner core \cite{GS1, GS2}
as the region above $R$  = 1050 km made of Ni-poor
$\delta$Fe(Ni) mush whose interstices have fluid in them. The layer between
870 and 1050 km is made of both Ni-rich $\gamma$Fe(Ni) and Ni-poor
$\delta$Fe(Ni) crystals and  it belongs to the slightly less isotropic
regime in 
\cite{GS1}. The innermost core is made of Ni-rich
$\gamma$Fe(Ni) crystals and it includes the more anisotropic regime in
\cite{GS1}. On the other hand, there is already seismic
analysis giving strong support for a hypothesis of fine-scale
structure within the top 300 km of the inner core \cite{VB}, and
describing the increase of anisotropy with depth \cite{SH}. The
possible hemispheric asymmetry could easily
follow if the Ni/Fe ratio is slightly variable locally.
 
\begin{figure}
\center
\epsfig{file=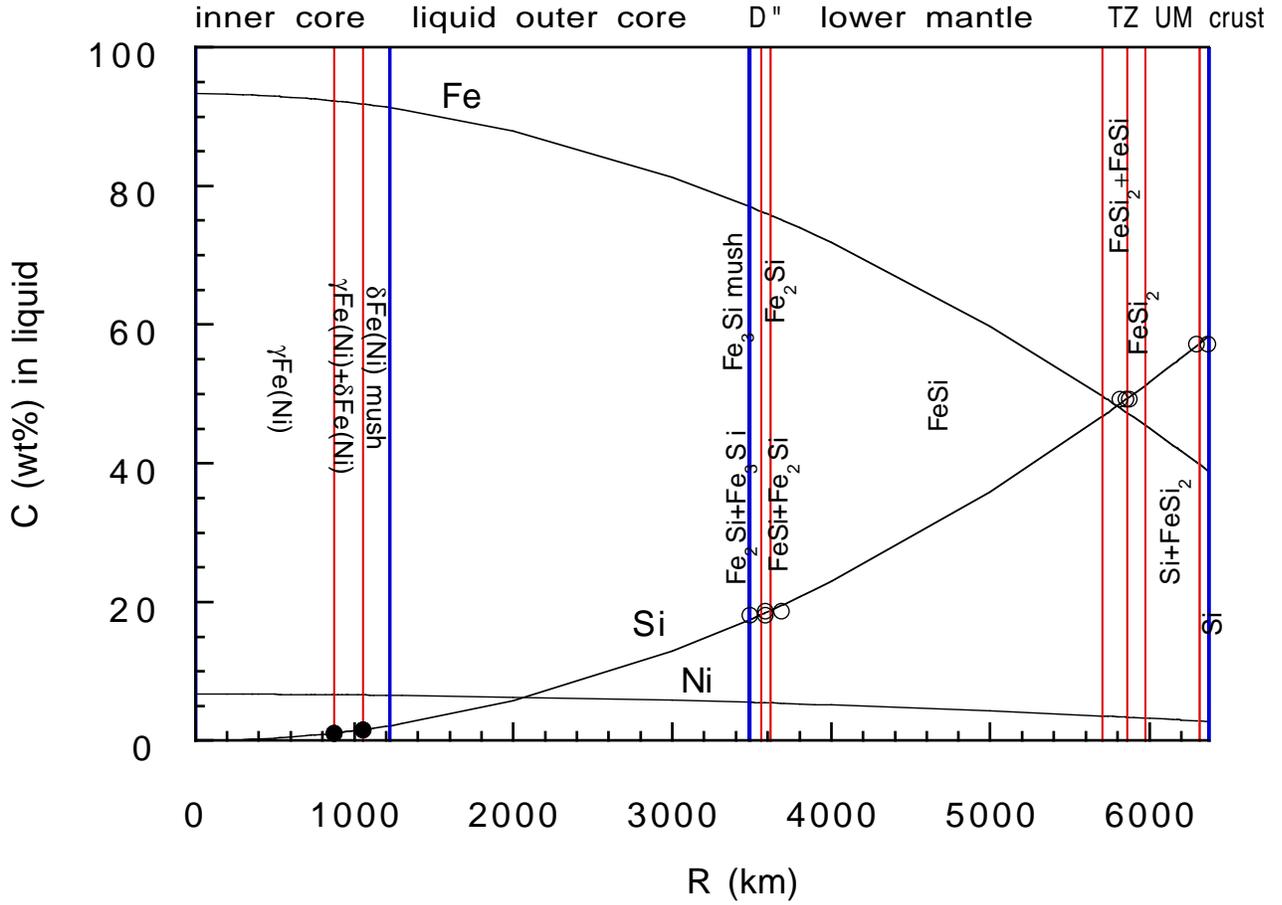,width=17cm}
\flushleft
\caption{{\small
\label{Fig. 4.} Fluid concentrations $C$ as a function of the radius
at the solidification  
front. The fit is $C_{Si} = a R^2$ with the points used in the fit
marked as open circles.  
This best fit was obtained with Ni/(Ni+Fe) = 0.067, this allows the
calculation  
of the other elements' concentrations $C_{Fe}$ and $C_{Ni}$. The
vertical lines separate  
the layers found in this model and their chemical structure. The two
solid dots  
indicate the Si-concentrations and radii of the structure change predicted by 
this model inside the inner core. The fronts at the present time are marked 
with bolder lines.
}}
\end{figure}

This model gives the (Si, Ni, Fe) fluid concentrations at the inner
core surface to be (0.021, 0.066, 0.913), and at the core-mantle
boundary (0.174, 0.055, 0.770).  The thickness of the D" layer is
determined by the fit to be 132 km which agrees very well with seismic
measurements \cite{VB}.  Its uppermost region is made of a
cotectic mixture of FeSi and Fe$_2$Si and below it is a layer of Fe$_2$Si,
but the location of their boundary is not available from the
model. The core-mantle boundary is located 72 km below the start of
the next layer, cotectically crystallizing Fe$_2$Si and Fe$_3$Si whose
thickness is not known, but is unlikely to be very wide, so that at
the present time the front is probably crystallizing Fe$_3$Si mush as
schematically shown in Fig. 3b.
 
\section{DISCUSSION AND CONCLUSIONS}
\setcounter{equation}{0}
\label{sec:tvs}

The silicate-rich outer parts and iron-rich inner parts of the planets
are the product of multicomponent solidification. Only partial
differentiation of the heavy and light elements needs to have
occurred. It would have been enough that gravity with convection
initially stratified the fluid in a continuous manner. Here this
stratification has been shown to depend on the second power of the
radial distance.

The Earth, if modeled by only the three components Fe, Si and Ni, with
the Ni/(Ni+Fe) ratio at the solidification front having a constant
value 0.067, would generate 13 chemically different layers just via
the process of three-component solidification. The layers are shown in
Table 1, but omitted from the Table is that each layer after the
outermost and innermost would first grow in the interstices of the
previous mush. This could reduce the sharpness of the
boundaries. Since the layers in multicomponent solidification are knit
together by mush growth the thermal boundary layers, expected to occur
for sharp chemical layering, would be absent. In addition, close to
the middle of the one-component phases the solid fraction of the
primary phase reaches its maximum, and this might have been observed
in some measurements about 1000 km above the CMB \cite{vK}. Here it
is estimated to occur at about $R$ = 4720 km, 1237 km above the
CMB.
 
The phase diagram lets one estimate the fluid concentrations where
each  cotectic layer first appears without the previous mush (for a
constant Ni/Fe ratio). Their radial locations have been obtained by
making an identification to seismic discontinuities.  The model gives
estimates for the fluid concentrations anywhere in the liquid outer
core and especially at the inner core and core-mantle boundaries. In
the outer core the Si-concentration is on average about 11 wt$\%$ which
agrees  very well with the density expectations based on seismic
measurements \cite{Bir}.

Similar abundances occur in the Sun.  The Si/Fe ratio in the Sun is
0.561 as calculated using data in \cite{BP}. In
this Earth model Si/Fe is 0.577 if averaged without any density
profile correction, or 0.520 if a very simple radial density profile
correction is employed, both rather close of the solar value.

This model gives a new, very  satisfactory account of the physical and
chemical basis of the well-documented seismic layered structure of the
Earth. The multitude of elements and chemicals in the real Earth could
possibly increase the number of layers, but it is not likely that any
of the layers described  in this paper disappear, since they follow
directly from the binary system of Fe-Si except in the inner core
where they follow from the binary system Fe-Ni. The chemical content of
the real Earth is, of course, enriched by the presence of the other elements.
This model considers
only the initial solidification at the liquidus surface and does not
take into account any later events like late impacts or the recycling
of the Earth's crust into the mantle at subduction zones, which affect
the chemical structure of the topmost layers.

 The seismic discontinuities at 400 and 670 km depth are identified
 here with chemical changes, not pressure-induced structural
 changes. The transition zone has here two chemically different
 layers: the upper one corresponds to FeSi$_2$ while the lower one has a
 mixture of FeSi$_2$+FeSi. The real chemical structure in these upper
 layers is known to be much modified by the presence of many more
 elements and the substantial recycling and recrystallisation of the
 materials. The model, however, agrees with the chemical type of the
 lower mantle: generally assumed to be dominantly (Mg,Fe)SiO$_3$ with
 perovskite structure. This corresponds to FeSi in this simplified
 model with only three components. 
Further inclusion of
 the two next commonest elements, namely O and Mg, would agree fully
 with our previous conceptions of the chemical compounds dominating
 the lower mantle.  Additional lighter elements are, of course, needed to
account for the mantle density in the real Earth, and their solidification is
 expected to have occurred in the interstices of the first solidified
 dominating compounds, thus generating also anisotropy. 
  
This model sheds light on the physical mechanism behind some seismic
observations in the middle of the lower mantle. The solid fraction of
FeSi-mush reaches its maximum at depth of about 1650 km (dashed line
in Fig. 2). It would act as a barrier of some strength to material
flow although there is no change in chemical composition across it,
only in the relative abundance of FeSi. A seismic low-velocity layer
\cite{KH} has been observed to occur at depths
1400 - 1600 km. Structural and chemical heterogeneity \cite{vK}
has been suggested to occur below this barrier at
depths 1700 - 2300 km. This heterogeneity is probably a consequence of
the appearance of the cotectically crystallized FeSi and Fe$_2$Si in the
interstices of the FeSi mush.

The D" layer has previously been anticipated to be chemically
different from the lower mantle by some authors \cite{SG}
even if its structure has not yet been agreed upon. The observed
40-60 km thick ridge-like ultra low velocity zone \cite{NH}
is probably due to the uneven \cite{Ai1} gupeiite
(Fe$_3$Si) mushy layer surface.

The crystal structure of Fe at high pressures is supposed to change
from $\delta$ to $\epsilon$. There are not adequate measurements to
indicate if this
happens also for Fe(Ni) at the inner core pressures and at the
liquidus temperature. But even if both structures $\gamma$ and $\delta$
change to be
$\epsilon$ at high pressures, the phases would be different due to their
different Ni-concentrations: the deepest inner core is rich in Ni, the
outermost mushy layer is  poor in Ni and the layer between is a
heterogeneous mixture of Ni-rich and Ni-poor Fe.

\vskip 20pt
{\centerline{\bf{ACKNOWLEDGMENTS}}}
\vspace{.25cm}
\noindent
I thank Michael Carpenter, John Hudson and Dan McKenzie for discussions. 
The main results of this work were presented as part of a poster "Ternary
solidification and beyond" at the 10th Anniversary Symposium of the
Daphne Jackson Trust on 21.1.2002 at the University of Surrey, Guildford,
Surrey.


\begin{table}

\begin{center}
\begin{tabular}{|c|c|c|c|}
\hline Depth (km) & Radius (km) &  Layer & Chemical compounds \\
\hline
(0)-59 &      (6371)-6312 & crust &            Si \\
59-(400) &    6312-(5971) & upper mantle  &    Si+FeSi$_2$ \\
(400)-513 &   (5971)-5858 & upper transition zone &  FeSi$_2$ \\
513-(670) &   5858-(5701) & lower transition zone &  FeSi$_2$+FeSi \\
(670)-2756 &  (5701)-3615 & lower mantle   &   FeSi \\
2756--    &  3615--     &  upper D" layer &   FeSi+Fe$_2$Si \\  
--2816 & --3555 &  middle D" layer &  Fe$_2$Si \\              
2816-- & 3555-- &     lower D" layer &   Fe$_2$Si+Fe$_3$Si \\
--(2888) & --(3483) &    lowest D" layer &  Fe$_3$Si-mush \\
(2888)-(5150)& (3483)-(1221) & outer core &  Fe+Ni+Si solution \\              
Inner core:  &  &  &  \\                               
(0)-171 & (1221)-1050 & outermost inner core & Ni-poor $\delta$Fe(Ni) mush\\
171-351 &  1050-870 & middle inner core & $\delta$Fe(Ni) + $\gamma$Fe(Ni)\\
351-(1221) & 870-(0) & innermost inner core & Ni-rich $\gamma$Fe(Ni)\\

\hline
\end{tabular}
\end{center}
\caption{Earth's layering. Numbers without brackets come
from this model.} 
\label{tab-sky-sa2}
\end{table}

\vskip 3cm

\newpage


\begin{thebibliography}{99}


\bibitem{Ag} Agee, C. B., 1993. Introduction to the special section on
magma oceans. {\it J. Geophys. Res.}, {\bf 98}, 5317.

\bibitem{Ai1} Aitta, A., Huppert, H. E. and Worster, M. G.,
2001a. Diffusion--controlled solidification of a ternary melt from a
cooled boundary.  {\it J. Fluid  Mech.} {\bf 432}, 201--217.

\bibitem{Ai2} Aitta, A., Huppert, H. E.  and Worster, M. G.,
2001b. Solidification in ternary systems. in {\it Interactive dynamics of
convection and solidification}  pp. 113--122, eds. Ehrhard, P., Riley,
D. S.  and Steen, P. H., Kluwer, Dordrecht.

\bibitem{ASM} ASM Handbook Vol. 3 {\it Alloy Phase Diagrams},
1992. ed. Baker, H., ASM International, Metals Park, Ohio.

\bibitem{Bi} Birch, F., 1952. Elasticity and constitution of the
Earth's interior. {\it J. Geophys. Res.}, {\bf 57}, 227--286.

\bibitem{Bir} Birch, F., 1964. Density and Composition of Mantle and
Core. {\it J. Geophys. Res.}, {\bf 69}, 4377--4388.

\bibitem{BP} Bahcall, J. N.  and Pinsonneault, M. H., 1995. Solar
models with helium and heavy-element diffusion. {\it Rev. Mod. Phys.}
{\bf 67}, 781--808.

\bibitem{Br} Braginsky, S. I., 1963. Structure of the F layer and
reasons for convection in the Earth's core. {\it Dokl. Akad. Nauk. SSSR}
(Earth Science) {\bf 149}, 8--10.

\bibitem{DW} Deuss, A.  and Woodhouse J., 2001. Seismic observations
of splitting of  the mid-transition zone discontinuity in Earth's
mantle. {\it Science} {\bf 294}, 354--357. 

\bibitem{DA} Dziewonski, A. M. and  Anderson, D. L., 1981. Preliminary
reference  Earth  model. {\it Phys.  Earth Planet. Inter.} {\bf 25},  297--356.

\bibitem{GS1} Garcia, R.  and Souriau, A., 2000. Inner core anisotropy
and heterogeneity level. {\it Geophys. Res. Lett.} {\bf 27}, 3121--3124.

\bibitem{GS2} Garcia, R.  and Souriau, A., 2001. Correction to "Inner
core anisotropy and heterogeneity level" by R. Garcia and
A. Souriau. {\it Geophys. Res. Lett.} {\bf 28}, 85--86.

\bibitem{KH} Kaneshima, S.  and Helffrich, G., 1999. Dipping
low-velocity layer in the mid-lower mantle: evidence for geochemical
heterogeneity.  {\it Science}  {\bf 283},  1888--1891.

\bibitem{MS} Masters, T. G.  and Shearer, P. M., 1995. Seismic models
of the Earth: Elastic and Anelastic. {\it Global Earth Physics:
Handbook of Physical Constants}  p. 94, ed. Ahrens, T. J. AGU,
Washington. 

\bibitem{NH} Ni, S.  and Helmberger, D. V., 2001. Probing an ultra-low
velocity zone at the core mantle boundary with P and S waves. {\it
Geophys. Res. Lett.} {\bf 28}, 2345--2348.

\bibitem{Po} Poirier, J.-P. {\it Introduction to the Physics of the Earth's
interior, 2nd Ed.} (Cambridge University Press, Cambridge 2000).

\bibitem{RR} Raynor, G. V.  and Rivlin, V. G., 1985. Critical
evaluation of constitution of cobalt-iron-silicon and
iron-nickel-silicon alloys. {\it Intern. Metals Rev.} {\bf 30}, 181--208.

\bibitem{SG} Sidorin, I.  and Gurnis, M., 1998. Geodynamically
consistent seismic velocity predictions at the base of the mantle.
{\it Geodynamics}  {\bf 28}, 209--230.

\bibitem{SH} Song, X.  and Helmberger, D. V., 1998. Seismic evidence
for an inner core transition zone. {\it Science}  {\bf 282}, 924--927.

\bibitem{vK} van der Hilst, R.  and Karason, H., 1999. Compositional
heterogeneity in the bottom 1000 kilometers of Earth's mantle: towards
a hybrid convection model. {\it Science}  {\bf 283}, 1885--1888.
 
\bibitem{VB} Vidale, J. E.  and Benz, H. M., 1993. Seismological
mapping of fine structure near the base of the Earth's mantle. {\it
Nature}  {\bf 361}, 529--532.

\bibitem{VE} Vidale, J. E.  and Earle, P. S., 2000. Fine-scale
heterogeneity in the Earth's inner core. {\it Nature} {\bf 404}, 273--275. 

\end{thebibliography}
\end{document}